\documentclass[12pt]{article}
\usepackage[cp1251]{inputenc}

\usepackage{amsfonts,amssymb,amsthm,mathtools}
\usepackage{amsmath}
\usepackage{icomma}
\usepackage{latexsym}
\usepackage{graphicx}
\usepackage{subfigure}
\usepackage {bm}

\usepackage{geometry} 
\begin{document}

\begin{center}
	\textbf{THE LACK OF VACUUM POLARIZATION IN QUANTUM ELECTRODYNAMICS WITH SPINORS IN FERMION EQUATIONS}
\end{center}

\begin{center}
	
	{V.~P.~Neznamov\footnote{vpneznamov@mail.ru, vpneznamov@vniief.ru}}\\
		
		\hfil
		{\it \mbox{	Russian Federal Nuclear Center--All-Russian Research Institute of Experimental Physics},  Mira pr., 37, Sarov, 607188, Russia \\
		National Research Nuclear University MEPhI, Moscow, 115409, Russia} \\
	\end{center}

%\maketitle

%\begin{history}
%\received{Day Month Year}
%\revised{Day Month Year}
%\end{history}

%%%%%%%%%%%%%%%%%%%%%%%%%%%%%%%%%%%%%%%%%%%%%%%%%%%%%%%%%%%%%%%%%%%%%%%%%%%%%%%%%%%%
\begin{abstract}
\noindent
\footnotesize{In this paper, the versions of quantum electrodynamics (QED) with spinors in fermion 
equations are briefly examined. In the new variants of the theory, the 
concept of vacuum polarization is unnecessary. The new content of fermion vacuum (without the Dirac sea) in the examined versions of QED leads to new physical consequences, part of which can be 
tested experimentally in the future.}\\



\end{abstract}

%\tableofcontents

\section{Introduction}	

In the standard quantum electrodynamics (QED), the Dirac equation with the 
bispinor wave function is used to describe fermion states \cite{bib1}. The Dirac equation has solutions with positive and negative energies. As a rule, the physical vacuum of the Dirac equation is described in terms of fully occupied states with negative energy (the Dirac sea).

In the Dirac vacuum, virtual creation and annihilation of electron-positron 
pairs is assumed. It is supposed that any electrical charge is surrounded 
by a cloud of virtual electron-positron pairs. It leads to decrease in initial charge, to reduce the effective Coulomb potential and to influence on 
the observed physical effects. This phenomenon is called vacuum 
polarization.

At electron scattering in the external electromagnetic field, the Feynman diagram, Fig.1, is considered for a manifestation of vacuum polarization in 
the lowest-order of the perturbation theory.

%%%%%%%%%%%%%%%%%%%%%%%%%%%%FIGURE 1%%%%%%%%%%%%%%%%%%%%%%%%%%%%%%%%%%%%%%%%%

\begin{figure}[h!]
	\center{\includegraphics[width=0.4\linewidth]{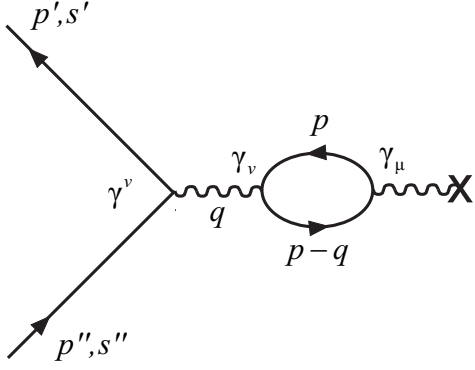}}
	\caption{The Feynman diagram for vacuum polarization in the standard QED.}
	\label{ris:Fig.1}
\end{figure}

%%%%%%%%%%%%%%%%%%%%%%%%%%%%%%%%%%%%%%%%%%%%%%%%%%%%%%%%%%%%%%%%%%%%%%%%%%%%%%

The accounting of the diagram, Fig. 1, contributes to the Lamb shift of the 
atomic levels. For hydrogen atom, this contribution to the level shift of 
$2S_{1/2} -2P_{1 / 2} $ is 27MHz.

The motion of fermions in the quantum theory can be described by equations 
with spinor functions. We will consider the two versions: the 
Foldy-Wouthuysen (FW) representation \cite{bib2} and the 
representation with Klein-Gordon (KG)-type equation for fermions \cite{bib3}. For these representations, we developed the 
formalisms of quantum electrodynamics (QED) $_{FW\, }$, (QED)$_{KG\, }$ and 
calculated some physical effects \cite{bib4}, \cite{bib5}.

In the lowest order of the perturbation theory, the Coulomb cross-sections 
of an electron scattering, the scattering of an electron on a proton, the 
Compton effect, annihilation of an electron-positron pair are calculated. 
The self-energy of an electron, the anomalous magnetic moment of an 
electron, the Lamb shift of atomic energy levels are also calculated. The 
final results completely agree with the appropriate results in the standard 
QED with Dirac equation \cite{bib4}, \cite{bib5}.

The following is new in (QED) $_{FW\, }$ and (QED) $_{KG}$:

\begin{itemize}
	\item[(i)] The equation for electrons does not relate to the equation for positrons. These equations differ from each other by the signs of electrical charges and by the signs in front of the mass terms.
	\item[(ii)] In each of the equations, there is no relationship between the solutions with positive and negative energies.
	\item[(iii)] The content of the physical fermion vacuum varies.
\end{itemize}
The existence of the sea of solutions with negative energy (the Dirac sea), 
the processes of virtual creation and annihilation of electron-positron 
pairs, the concept of vacuum polarization seem to be excessive. The new 
content of the physical fermion vacuum leads to new physical effects.

Section 2 presents a brief review of the (QED) $_{FW\, }$ and (QED) $ _{KG}$. 
In Secs. 3 and 4 we discuss the (QED) $_{FW}$ and (QED) $_{KG\, }$ physical 
fermion vacuum and possible new physical effects. In the conclusion, we 
formulate the main conclusions of the paper.

In what follows, we use the system of units of $\hslash =c=1$ and the Minkowski 
space-time signature
\[
g_{\alpha \beta } =\mbox{diag}\left\{ {1,-1,-1,-1} \right\}.
\]

%%%%%%%%%%%%%%%%%%%%%%%%%%%%%%%%%%%%%%%%%%%%%%%%%%%%%%%%%%%%%%%%%%%%%%%%%%%%

\section{The formalism of QED with spinor equations for fermions}
The Dirac equation for an electron with mass $m$ interacting with the 
electromagnetic field can be written as
\begin{equation}
	\label{eq1}
	p^{0}\psi_{D} =H_{D} \psi_{D} =\left( {{\bm {\alpha }}\left( {{\rm 
				{\bf p}}-e{\rm {\bf A}}} \right)+\beta m+eA^{0}} \right)\psi_{D} ,
\end{equation}
where $H_{D} $ is the Dirac Hamiltonian, $p^{0}=i\left( {\partial 
	\mathord{\left/ {\vphantom {\partial {\partial t}}} \right. 
		\kern-\nulldelimiterspace} {\partial t}} \right),\,\,{\rm {\bf 
		p}}=-i\vec{{\nabla }}$; $A^{0}\left( {{\rm {\bf r}},t} \right),\,\,{\rm {\bf 
		A}}^{i}\left( {{\rm {\bf r}},t} \right)$ are electromagnetic potentials; 
$\alpha^{k},\,\beta $ are four-dimensional Dirac matrixes; $k=1,2,3$.

The bispinor $\psi_{D} $ has the form
\begin{equation}
	\label{eq2}
	\psi_{D} =\left( {{\begin{array}{*{20}c}
				{\varphi \left( {{\rm {\bf x}},t} \right)U_{S} } \hfill \\
				{\chi \left( {{\rm {\bf x}},t} \right)U_{S} } \hfill \\
	\end{array} }} \right).
\end{equation}

The following equation can be used for description of positrons:
\begin{equation}
	\label{eq3}
	p^{0}\psi_{D}^{C} =\left( {{\bm {\alpha }}\left( {{\rm {\bf p}}+e{\rm 
				{\bf A}}} \right)-\beta m-eA^{0}} \right)\psi_{D}^{C} .
\end{equation}

Here $\psi_{D}^{C}= \beta \Sigma _2 \psi_{D}^{*}$, $\Sigma_2 = \left( {{\begin{array}{*{20}c}
			{\sigma_2 \,\,\,\ 0 } \hfill \\
			{0 \,\,\,\ \sigma_2 } \hfill \\
\end{array} }} \right) $, $\psi_{D}^{*}$ is complex conjugate bispinor.

Equation (\ref{eq3}) differs from Eq. (\ref{eq1}) by the signs of charge and mass.

In the free case (without interaction), the Dirac equations (\ref{eq1}) and (\ref{eq3}) have the following normalized solutions with positive and negative energies:
\begin{equation}
	\label{eq4}
	\begin{array}{l}
		\psi_{D}^{\left( + \right)} \left( {{\rm {\bf x}},t;+m} \right)=\sqrt 
		{\dfrac{E+m}{2E}} \left( {\begin{array}{l}
				\,\,\,\,\,\,\,\,U_{S} \\ [10pt]
				\dfrac{{\rm {\bm \sigma \bf p}}}{E+m}U_{S} \\  [15pt]
		\end{array}} \right)e^{-iEt+i{\rm {\bf px}}}, \\ 
		\psi_{D}^{\left( - \right)} \left( {{\rm {\bf x}},t;+m} \right)=\sqrt 
		{\dfrac{E+m}{2E}} \left( {\begin{array}{l}
				\dfrac{{\rm {\bm \sigma \bf p}}}{E+m}U_{S} \\ [10pt]
				\,\,\,\,\,\,\,\,U_{S} \,\,\, \\ 
		\end{array}} \right)e^{iEt-i{\rm {\bf px}}}. \\ 
	\end{array}
\end{equation}
\begin{equation}
	\label{eq5}
	\begin{array}{l}
		\psi_{D}^{C\,\,\left( + \right)} \left( {{\rm {\bf x}},t;-m} \right)=\sqrt 
		{\dfrac{E-m}{2E}} \left( {\begin{array}{l}
				\,\,\,\,\,\,\,\,U_{S}^C \\ [10pt]
				\dfrac{{\rm {\bm \sigma \bf p}}}{E-m}U_{S}^C \\ [15pt]
		\end{array}} \right)e^{-iEt+i{\rm {\bf px}}}, \\ 
		\psi_{D}^{C\,\,\left( - \right)} \left( {{\rm {\bf x}},t;-m} \right)=\sqrt 
		{\dfrac{E-m}{2E}} \left( {\begin{array}{l}
				\dfrac{{\rm {\bm \sigma \bf p}}}{E-m}U_{S}^C \\ 
				\,\,\,\,\,\,\,\,U_{S}^C \,\,\, \\ [10pt]
		\end{array}} \right)e^{iEt-i{\rm {\bf px}}}. \\ 
	\end{array}
\end{equation}
In Eqs. (\ref{eq2}) -- (\ref{eq5}), $E=\sqrt {m^{2}+{\rm {\bf p}}^{2}} $; $\sigma^{k}$ are two-dimensional Pauli matrixes; $U_{S}^C = \sigma ^2 U_{S}$ are normalized Pauli spinors.

Solutions (\ref{eq4}) and (\ref{eq5}) were obtained by using matrices $\alpha^{k},\,\beta $ in the Dirac-Pauli representation. The similar solutions can be obtained with Dirac matrixes in the spinor representation widely used in the Standard Model. The QED with spinor equations for fermions and with the spinor representation of Dirac matrixes is presented in paper \cite{bib6} - (QED) $_{FW\, }$ and in Ref. \cite{bib7} - (QED) $_{KG}$. The final physical results in Refs. \cite{bib6} and \cite{bib7} coincide with the results in standard QED and with the results in Refs. \cite{bib4} and \cite{bib5} by using matrixes $\alpha^{k},\,\beta $ in the Dirac-Pauli representation.

%%%%%%%%%%%%%%%%%%%%%%%%%%%%%%%%%%%%%%%%%%%%%%%%%%%%%%%%%%%%%%%%%%%%%%%%%%%%%%

\subsection{QED in Foldy-Wouthuysen representation (QED) $_{\mathbf{FW}}$}

The Dirac equation in the FW representation can be obtained as a power 
series expansion in the electromagnetic coupling constant by using a number 
of unitary transformations $U_{FW} =\left( {1+\delta_{1} +\delta_{2} 
	+\delta_{3} +...} \right)U_{0} $ \cite{bib4}.

As the results, we obtain the following equations:
\begin{equation}
	\label{eq6}
	\begin{array}{l}
	p^{0}\psi_{FW} =H_{FW} \psi_{FW} =\left( {\beta E+eK_{1} \left( {+m,A^{\mu 
		}} \right)+e^{2}K_{2} \left( {+m,A^{\mu },A^{\nu }} \right)+} \right. \\[10pt]
	\left. {+ e^{3}K_{3} 
		\left( {+m,A^{\mu },A^{\nu },A^{\gamma }} \right)+...} \right)\psi_{FW} .
	\end{array}
\end{equation}
\begin{equation}
	\label{eq7}
	\begin{array}{l}
		p^{0}\psi_{FW}^{C} =H_{FW}^{C} \psi_{FW}^{C} =\left( {\beta E-eK_{1} 
		\left( {-m,A^{\mu }} \right)+e^{2}K_{2} \left( {-m,A^{\mu },A^{\nu }} 
		\right)- } \right.\\ [10pt]
	\left. { -e^{3}K_{3} \left( {-m,A^{\mu },A^{\nu },A^{\gamma }} \right)+...} 
	\right)\psi_{FW}^{C} .
	\end{array}
\end{equation}
In the free case, we have
\begin{equation}
	\label{eq8}
	p_{0} \psi_{FW}^{0} =\beta E\psi_{FW}^{0} ,
\end{equation}
where for the positive energy of $p_{0} =E$
\begin{equation}
	\label{eq9}
	\psi_{FW}^{0\,\,\left( + \right)} \left( {{\rm {\bf x}},t} \right)=\left( 
	{\begin{array}{l}
			U_{S} \\ 
			\,0 \\ 
	\end{array}} \right)e^{-iEt+i{\rm {\bf px}}},
\end{equation}
For the negative energy of $p_{0} =-E$, we have
\begin{equation}
	\label{eq10}
	\psi_{FW}^{0\,\,\left( - \right)} \left( {{\rm {\bf x}},t} \right)=\left( 
	{\begin{array}{l}
			\,0 \\ 
			\,U_{S} \\ 
	\end{array}} \right)e^{iEt-i{\rm {\bf px}}}.
\end{equation}
In case of interaction, Hamiltonians $H_{FW} ,H_{FW}^{C} $ in (\ref{eq6}) and (\ref{eq7}) are diagonal relative to the mixing of the upper and lower components of bispinor $\psi_{FW} $. Each of Eqs. (\ref{eq6}) and (\ref{eq7}) actually includes two independent equations with spinor wave functions $\sim U_{S} $.

In the equations of the FW-representation, masses of particles and 
antiparticles have the opposite signs \cite{bib4}. If we use 
solutions (\ref{eq6}) with positive energy and mass $\left( {+m} \right)$ for 
electrons, then for positrons, we should use solutions (\ref{eq7}) with positive energy and mass $\left( {-m} \right)$.

In Ref. \cite{bib4}, the formalism of quantum 
electrodynamics (QED) $_{FW\, }$ is developed and some of physical effects 
are calculated. The final calculation results coincide with the results in 
the standard QED.

In (QED) $_{FW\, }$, the concept of vacuum polarization is unnecessary.

%%%%%%%%%%%%%%%%%%%%%%%%%%%%%%%%%%%%%%%%%%%%%%%%%%%%%%%%%%%%%%%%%%%%%%%%%%%%%%%%%%%%%%%%
\subsection{QED with the spinor equation for fermions of Klein-Gordon type}

The self-conjugated equations for electrons and positrons with spinor wave 
functions are obtained in Refs. \cite{bib3} and \cite{bib5}. These equations are given by
\begin{equation}
	\label{eq11}
	\begin{array}{l}
	\left[ {\left( {p^{0}-eA^{0}} \right)^{2}-m^{2}-\left( {p^{0}-eA^{0}+m} 
		\right)^{1 /2}{\rm {\bm \sigma }}\left( {{\rm {\bf p}}-e{\rm 
				{\bf A}}} \right) } \times \right. \\ [10pt]
		\left. { \times \dfrac{1}{p^{0}-eA^{0}+m}{\rm {\bm \sigma }}\left( {{\rm 
				{\bf p}}-e{\rm {\bf A}}} \right)\left( {p^{0}-eA^{0}+m} \right)^{1 	/2}} 
	\right]\Phi =0,
\end{array}
\end{equation}
\begin{equation}
	\label{eq12}
	\begin{array}{l}
		\left[ {\left( {p^{0}+eA^{0}} \right)^{2}-m^{2}-\left( {p^{0}+eA^{0}-m} 
			\right)^{1 /2}{\rm {\bm \sigma }}\left( {{\rm {\bf p}}+e{\rm 
					{\bf A}}} \right)} \times \right. \\ [10pt]
		\left. {\times \dfrac{1}{p^{0}+eA^{0}-m}{\rm {\bm \sigma }}\left( {{\rm {\bf 	p}}+e{\rm {\bf A}}} \right)\left( {p^{0}-eA^{0}-m} \right)^{1 	/ 2}} 
		\right]\Phi^{C}=0. \\ 
	\end{array}
\end{equation}
In Eqs. (\ref{eq11}) and (\ref{eq12}), we can perform expansion in power of the charge $e$
\begin{equation}
	\label{eq13}
	\begin{array}{l}
	\left[ {p_{0}^{2} -{\rm {\bf p}}^{2}-m^{2} \mp eV_{1} \left( {\pm m,A^{\mu }} 
		\right)-e^{2}V_{2} \left( {\pm m,A^{\mu },A^{\nu }} \right) \mp } \right. \\ [10pt]
	\left. { \mp e^{3}V_{3} 
		\left( {\pm m,A^{\mu },A^{\nu },A^{\gamma }} \right)-...} \right]\Phi \left( 
	{\pm m,{\rm {\bf x}},t} \right)=0.
\end{array}
\end{equation}
In (\ref{eq13}), the upper signs before charge and mass correspond to Eq. (\ref{eq11}) for electron, the lower signs correspond to Eq. (\ref{eq12}) for positron. In Eqs. (\ref{eq11}) and (\ref{eq13}) $\Phi \left( {+m,{\rm {\bf x}},t} 
\right)=\Phi $, in Eqs. (\ref{eq12}) and (\ref{eq13}) $\Phi \left( {-m,{\rm {\bf x}},t} \right)=\Phi^{C}$.

The algorithm to determine the interaction operator of $V=\pm eV_{1} +e^{2}V_{2} 
\pm e^{3}V_{3} +...$ is given in Ref. \cite{bib5}.

In the free case, Eqs. (\ref{eq11}) and (\ref{eq12}) have the form of KG equations with spinor wave functions
\begin{equation}
	\label{eq14}
	\left( {p_{0}^{2} -{\rm {\bf p}}^{2}-m^{2}} \right)\Phi_{0} \left( {{\rm 
			{\bf x}},t} \right)=0.
\end{equation}
The orthonormal solutions $\Phi_{0}^{\left( \pm \right)} $ with positive 
and negative energies are given by
\begin{equation}
	\label{eq15}
	\Phi_{0}^{\left( \pm \right)} \left( {{\rm {\bf x}},t} 
	\right)=\frac{1}{\sqrt {2E} }e^{\mp iEt\pm i{\rm {\bf px}}}U_{S} .
\end{equation}
In Ref. \cite{bib5}, the formalism of quantum 
electrodynamics (QED) $_{KG\, }$ is developed and some of physical effects 
are calculated. As well as in the representation (FW), the final 
computational results coincide with the results in the standard QED.

In (QED) $_{KG\, }$, the solutions with positive and negative energies of 
fermions in Eq. (\ref{eq13}) are not connected with each other. In (QED) $_{KG}$ equations, the masses of particles and antiparticles have different signs. In 
(QED) $_{KG}$, the content of physical vacuum varies and the concept of 
vacuum polarization is unnecessary.

%%%%%%%%%%%%%%%%%%%%%%%%%%%%%%%%%%%%%%%%%%%%%%%%%%%%%%%%%%%%%%%%%%%%%%%%%%%

\section{Physical vacuum for fermions: the lack of vacuum polarization in QED with spinor equations for fermions}

In the standard QED with the Dirac equation, the fermion vacuum represents 
the continuum of fully occupied states with negative energies (the Dirac 
sea, see Fig.2)
%%%%%%%%%%%%FIGURE 2%%%%%%%%%%%%%%%%%%%%%%%%%%%%%%%%%%%%%%%%%%%%%%%%%%%%%%%

\begin{figure}[h!]
	\centerline{\includegraphics[width=0.8\linewidth]{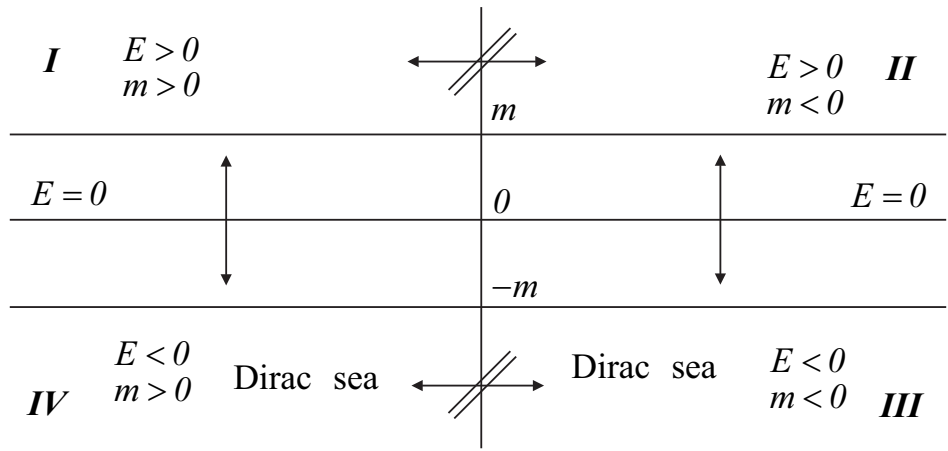}} 
	\caption{The physical vacuum of the Dirac equation.}
	\label{ris:Fig.2}
\end{figure}

%%%%%%%%%%%%%%%%%%%%%%%%%%%%%%%%%%%%%%%%%%%%%%%%%%%%%%%%%%%%%%%%%%%%%%%%%%%
In the standard QED, there exists the interaction of states with positive 
and negative energies. The solutions with different signs before the 
particle masses do not interact with each other. If we use quadrants I and 
IV of Fig. 2 with $m>0$, then the solutions with $m<0$ in quadrants II and 
III do not carry new physical information.

The holes in quadrant IV of Fig. 2 with $E<0$ represent the states of 
antiparticles in the standard QED. In theory, the possibility of spontaneous 
vacuum creation and annihilation of virtual particle-antiparticle pairs is 
admitted. As the result, the concept of vacuum polarization emerges. It is 
assumed that any charge is surrounded by a cloud of virtual 
particle-antiparticle pairs. It leads to efficient decrease in the value of 
the bare electrical charge that manifests itself in calculations of the Lamb 
shift of atomic energy levels.

Now, let us turn to the physical vacuum of the fermion equations with spinor 
functions (see Fig. 3).

%%%%%%%%%%%%%%FIGURE 3%%%%%%%%%%%%%%%%%%%%%%%%%%%%%%%%%%%%%%%%%%%%%%%%%%%%%%%%

\begin{figure}[h]
	\centerline{\includegraphics[width=0.8\linewidth]{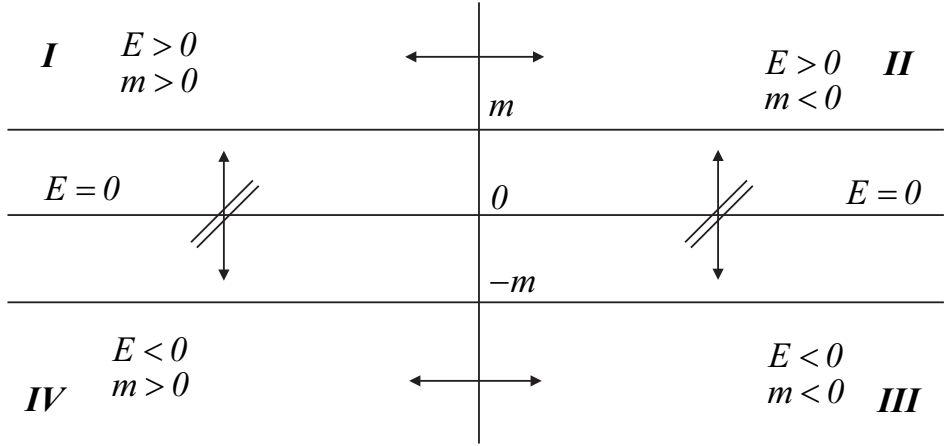}}
	\caption{The physical vacuum of fermion equations with spinor functions.}
	\label{ris:Fig.3}
\end{figure}

%%%%%%%%%%%%%%%%%%%%%%%%%%%%%%%%%%%%%%%%%%%%%%%%%%%%%%%%%%%%%%%%%%%%%%%%%%%%%%

Unlike the standard QED, there is no interaction between the states with 
positive and negative energies in the (QED) $_{FW\, }$ and (QED) $_{KG\, }$. 
In this connection, an introduction of the Dirac sea is unnecessary.

The physical fermion vacuum in (QED) $_{FW\, }$ and (QED) $_{KG\, }$, when 
using quadrants I, II, represents completely unoccupied states of particles 
with $E>0,m>0$ and completely unoccupied states of antiparticles with 
$E>0,m<0$. In equations of this theory, masses of particles and antiparticles should always 
have different signs.

While using quadrant I with $E>0,m>0$ for particles, the calculations of 
concrete physical effects in (QED) $_{FW\, }$ and (QED) $_{KG\, }$ are 
performed only with participation of intermediate (virtual) states with 
positive energy. For antiparticles of quadrant II with $E>0,m<0$, only 
virtual states with positive energy are also involved. In the theories under 
consideration, there is no need to take into account the process of creation 
and annihilation of virtual particle-antiparticle pairs. While using 
Eqs. (\ref{eq6}), (\ref{eq7}), (\ref{eq11}) and (\ref{eq12}) in (QED) $_{FW\, }$ and (QED )$_{KG}$, only the processes with real particles and antiparticles with opposite signs before their masses are taken into account. Only in this case, the interaction of the particles from the quadrant I with the antiparticles from the quadrant II is possible.

In Figs. 4 and 5, the Feynman diagrams in (QED) $_{FW}$, (QED) $_{KG}$, related 
to the self-energy function of a photon, and equivalent to the diagram in 
Fig. 1 in standard QED are presented in case of electron scattering in the 
external electromagnetic field.

%%%%%%%%%%%%%%%%%%%%%%%%FIGURE 4%%%%%%%%%%%%%%%%%%%%%%%%%%%%%%%%%%%%%%%%%%%

\begin{figure}[h]
	\begin{minipage}[h]{0.32\linewidth}
		\center{\includegraphics[width=1\linewidth]{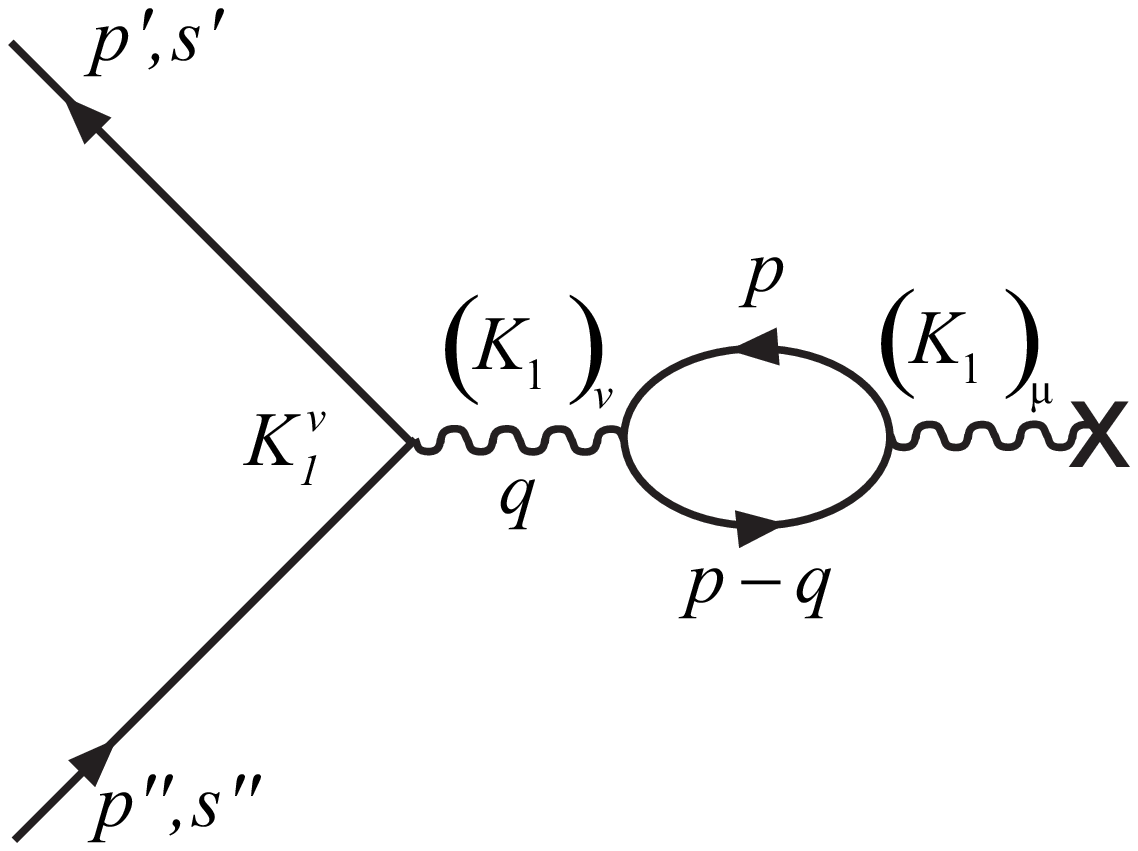} \\ a)}
	\end{minipage}
	\hfill
	\begin{minipage}[h]{0.32\linewidth}
		\center{\includegraphics[width=1\linewidth]{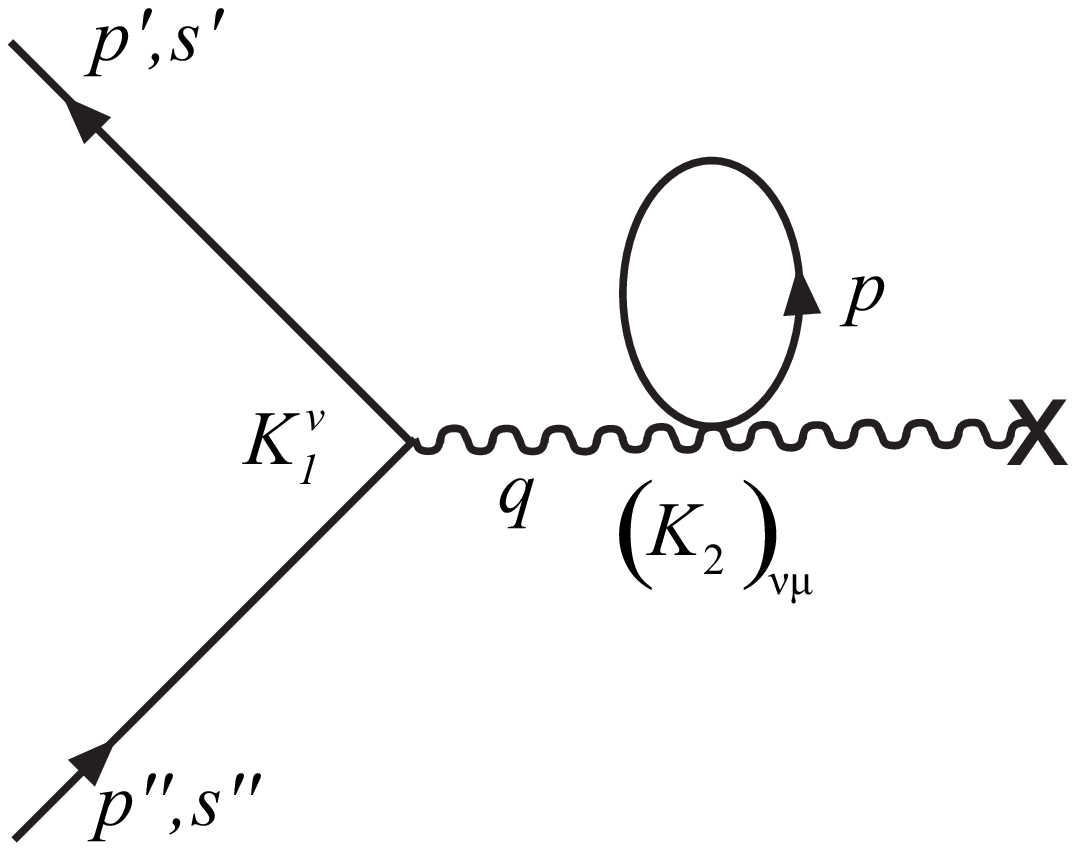} \\ b)}
	\end{minipage}
	\hfill
	\begin{minipage}[h]{0.32\linewidth}
		\center{\includegraphics[width=1\linewidth]{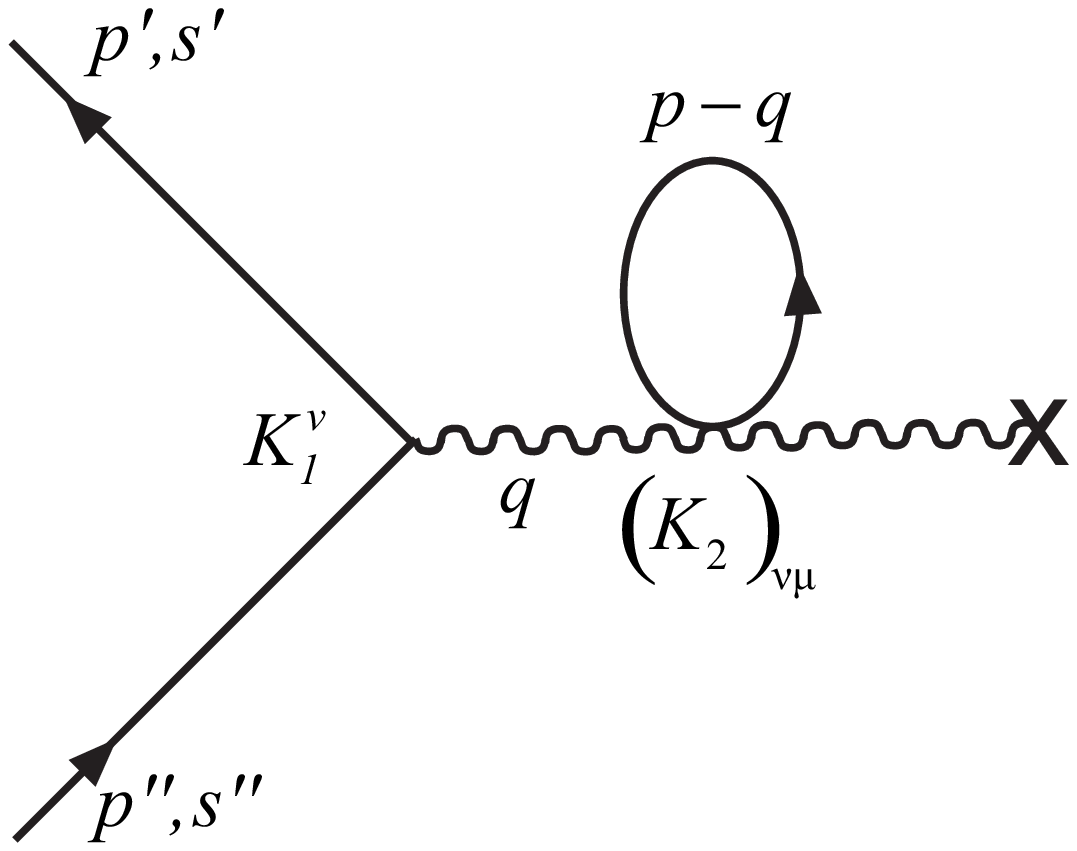} \\ c)}
	\end{minipage}
	\caption{The Feynman diagrams in (QED)$_{FW}$.}
	\label{Fig.4}
\end{figure}

%%%%%%%%%%%%%%%%%%%%%%%%%%%%%%%%%%%%%%%%%%%%%%%%%%%%%%%%%%%%%%%%%%%%%%%%%%%

%%%%%%%%%%%%%%%%%%%%%%%%FIGURE 5%%%%%%%%%%%%%%%%%%%%%%%%%%%%%%%%%%%%%%%%%%%

\begin{figure}[h]
	\begin{minipage}[h]{0.32\linewidth}
		\center{\includegraphics[width=1\linewidth]{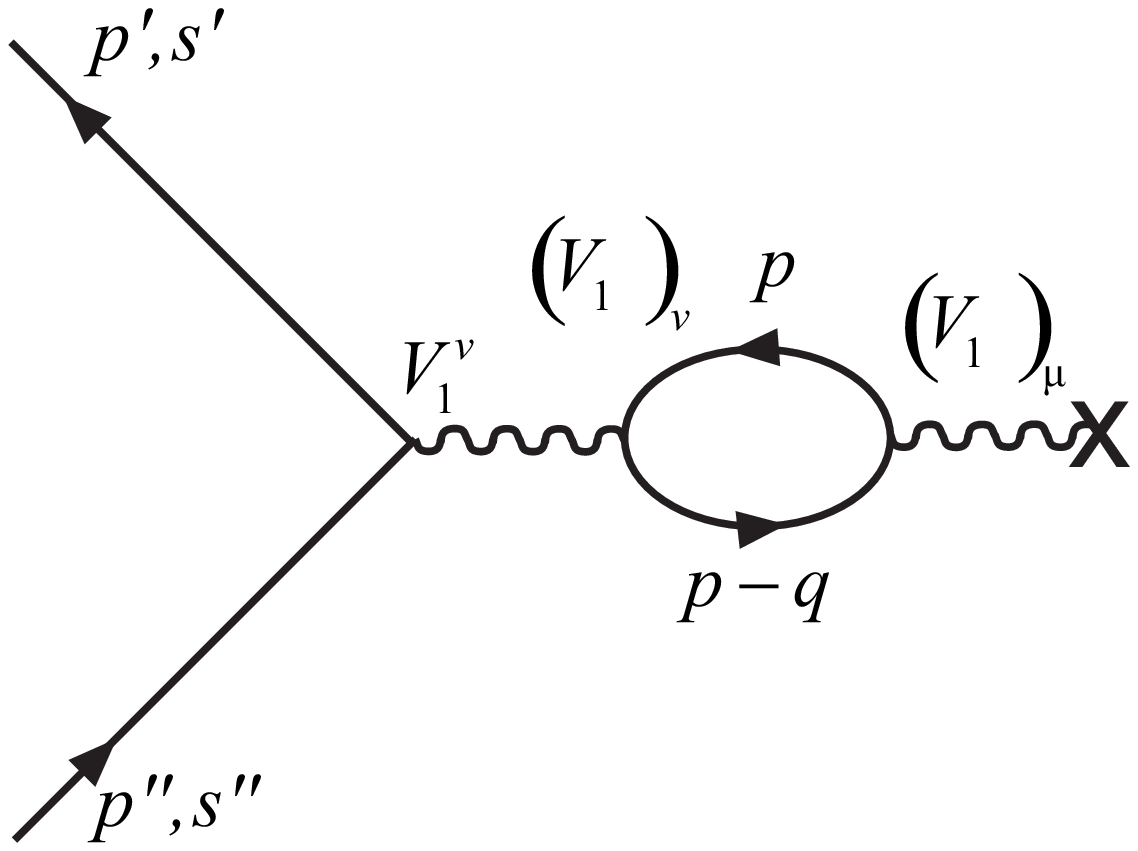} \\ a)}
	\end{minipage}
	\hfill
	\begin{minipage}[h]{0.32\linewidth}
		\center{\includegraphics[width=1\linewidth]{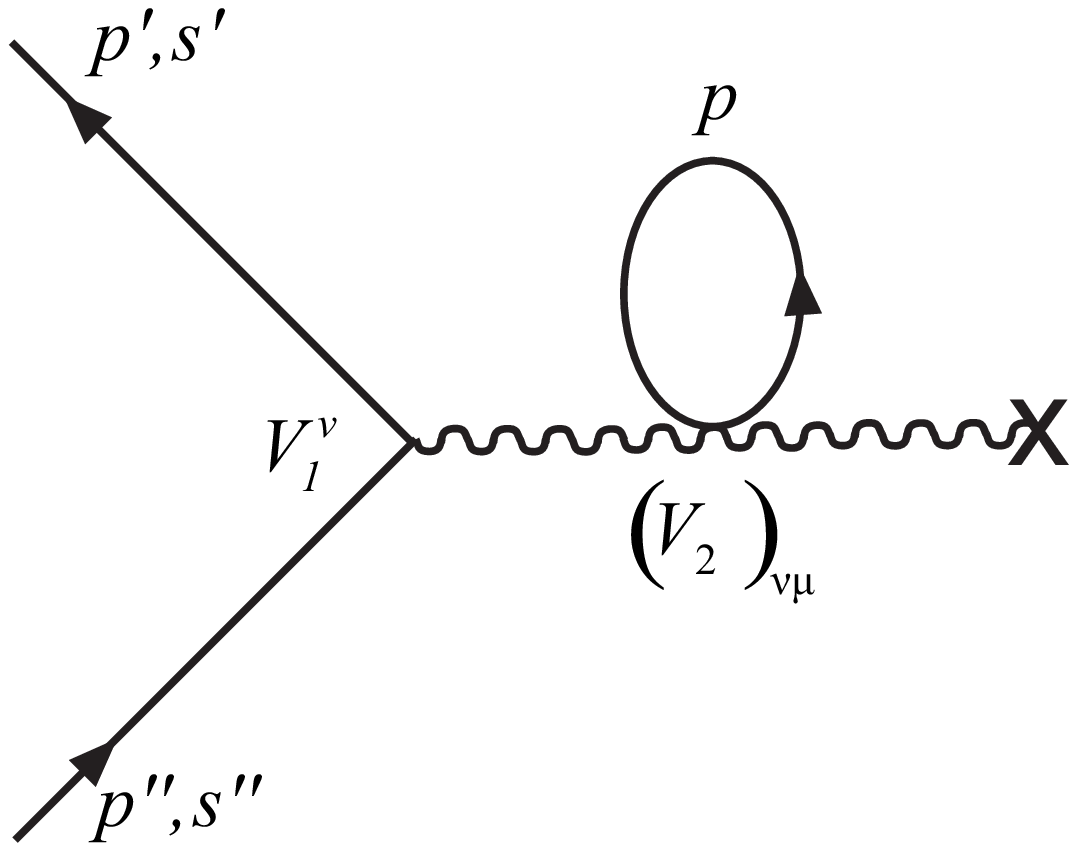} \\ b)}
	\end{minipage}
	\hfill
	\begin{minipage}[h]{0.32\linewidth}
		\center{\includegraphics[width=1\linewidth]{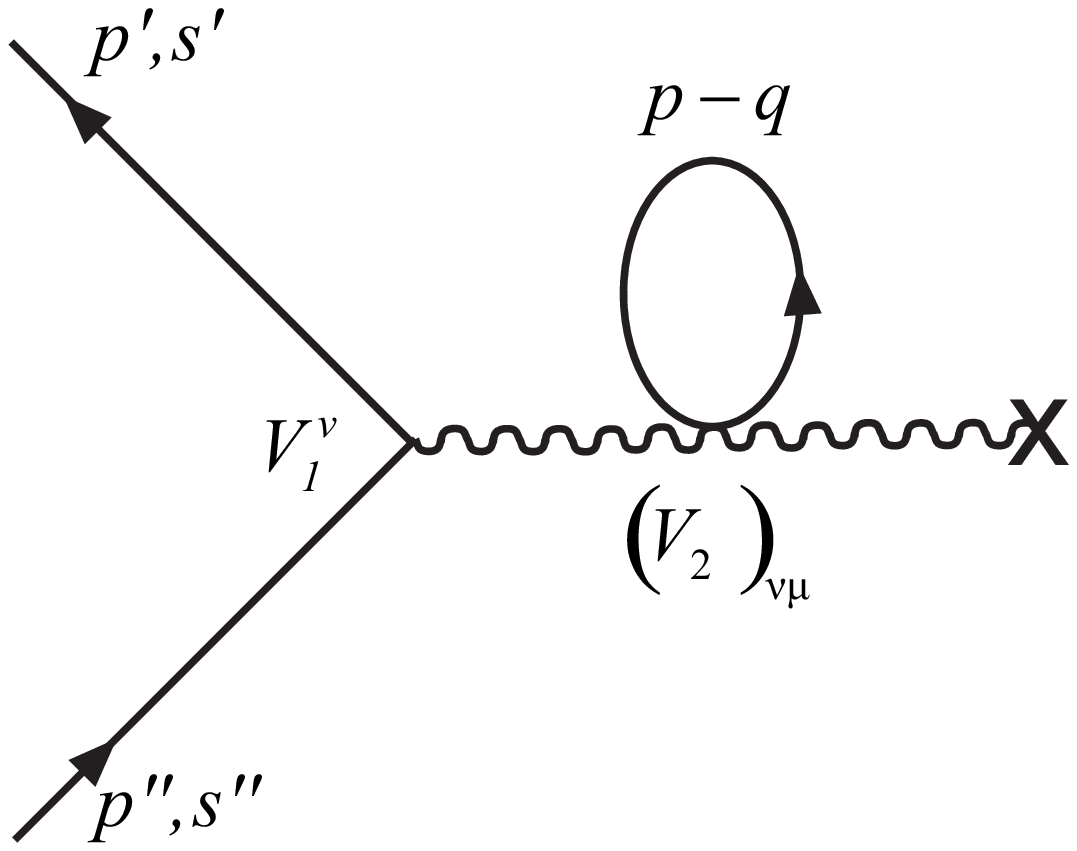} \\ c)}
	\end{minipage}
	\caption{The Feynman diagrams in (QED)$_{KG.}$}
	\label{Fig.5}
\end{figure}

%%%%%%%%%%%%%%%%%%%%%%%%%%%%%%%%%%%%%%%%%%%%%%%%%%%%%%%%%%%%%%%%%%%%%%%%%%%

The final results of the computations of the diagram in Fig.1 in the 
standard QED coincide with the computational results of the diagrams in Fig. 
4 in (QED) $_{FW\, }$ and the computational results of the diagrams in Fig. 5 
in (QED) $_{KG}$.

However, while diagrams in Figs. 4(a) and 5(a) can be described in terms of creation and 
annihilation of virtual electron-positron pairs, diagrams in Figs. 4(b), 4(c) and 5(b), 5(c) 
cannot be interpreted in this way. Let us note again that the all physically 
observed effects for particles and antiparticles are described by using only 
the intermediate states with positive energy.

It follows from the above-mentioned that there are no processes of creation 
and annihilation of virtual fermion ''particle-antiparticle'' pairs in 
(QED) $_{FW\, }$ and (QED) $_{KG}$. In these theories, the concept of vacuum 
polarization is unnecessary.

%%%%%%%%%%%%%%%%%%%%%%%%%%%%%%%%%%%%%%%%%%%%%%%%%%%%%%%%%%%%%%%%%%%%%%%%%%%%%

\section{Discussion}
The new state of the fermion vacuum in (QED)$_{FW\, }$ and (QED)$_{KG\, }$ 
leads to new physical consequences. Some of them can be tested 
experimentally in the future:

\begin{itemize}
	\item[(i)] In (QED) $_{FW\, }$ and (QED) $_{KG}$, there is no Zitterbevegung of 
fermion coordinates. This fact, associated with the absence of virtual 
interaction between the states of the fermions with positive and negative 
energies, was already mentioned in the first article of Foldy-Wouthuysen \cite{bib2}.

	\item [(ii)]For the same reasons in (QED) $_{FW\, }$ and (QED) $_{KG}$, there is no Klein paradox \cite{bib8} (see App. A and Ref. \cite{bib9}).

	\item[(iii)] In (QED) $_{FW\, }$ and (QED) $_{KG}$, there is no effect of vacuum 
creation of fermion particle-antiparticle pairs in strong electromagnetic 
fields. The Schwinger effect, i.e. vacuum creation of pairs in a strong 
homogeneous electrical field, is also absent \cite{bib10}.

	\item[(iv)] Because of absence of the Dirac sea in (QED) $_{FW\, }$ and (QED) $_{KG}$, the effect of vacuum creation of two electron-positron pairs is also absent at the achievement of the value of the nucleus charge of $Z=Z_{cr} \simeq 170$ for $1S_{1 / 2} $-state of the hydrogen-like atom \cite{bib11}. In the theories under consideration, a minimal 
possible value of energy $E=0$ will be achieved for $1S_{1 / 2} $-state at $Z_{1S_{1 /2} } \simeq 145$ already \cite{bib11}. With the following 
increase in $Z$, the level $1S_{1/ 2} $ disappears. The next level of $2P_{1 
	/2} $ disappears at $Z_{2P_{1 / 2} } \simeq 170$. The values of $Z_{1S_{1 
		/ 2} } , Z_{2P_{1 /2} } $ etc. depend on the model of the finite 
size of an atomic nucleus \cite{bib12}.

	\item[(v)] In quantum gravitational theories, similar to (QED) $_{FW\, }$ and 
(QED) $_{KG}$, there is no effect vacuum creation of fermion 
particle-antiparticle pairs. In this case, evaporation of black holes is 
possible only at the expense of vacuum creation of boson pairs.

	\item[(vi)] The analysis of the equations with spinor wave functions in the Coulomb repulsive field in (QED) $_{KG\, }$ has shown the availability of the impenetrable potential barrier in the effective potential with a radius proportional to the classical fermion radius and inversely proportional to the fermion energy (at $E\gg mc^{2})$ \cite{bib3}. The 
existence of the impenetrable barrier does not contradict the results of the 
experiments in probing the internal structure of an electron and has no 
effect on the cross section of the Coulomb scattering of electrons in the 
lowest-order of the perturbation theory.

	\item[(vii)] In equations of (QED) $_{FW}$ and (QED) $_{KG}$, masses of particles and antiparticles have the different signs \cite{bib4, bib5}. For the first time author shown this in 1989 (see Refs. \cite{bib13} and \cite{bib14}). Later, the other researchers came to the same conclusion (see, for example, Refs. \cite{bib15} and \cite{bib16}). In this paper, we do not establish linkage between different signs before masses of particles and antiparticles with problems of gravitation and antigravitation.
\end{itemize}

\section{Conclusion}
In versions of quantum electrodynamics with spinors in fermion equations, 
the concept of vacuum polarization is unnecessary.

The new content of the fermion vacuum (without the Dirac sea) leads to new 
physical consequences, part of which can be experimentally tested in the 
future.

%%%%%%%%%%%%%%%%%%%%%%%%%%%%%%%%%%%%%%%%%%%%%%%%%%%%%%%%%%%%%%%%%%%%%%%%%%%%%

\section*{Acknowledgment}
The author expresses his gratitude to A.L.Novoselova for essential technical 
support in preparation of the paper.

%%%%%%%%%%%%%%%%%%%%%%%%%%%%%%%%%%%%%%%%%%%%%%%%%%%%%%%%%%%%%%%%%%%%%%%%%%%%%

%\appendix

\section*{Appendix A. The Foldy-Wouthuysen representation: the scattering on a step 	potential}

%%%%%%%%%%%%%%%%%%%%%%%%%%%%%%%%%%%FIGURE APPENDIX%%%%%%%%%%%%%%%%%%%%%%%%%%%%%%%%

\begin{figure}[h!]
	\center{\includegraphics[width=0.5\linewidth]{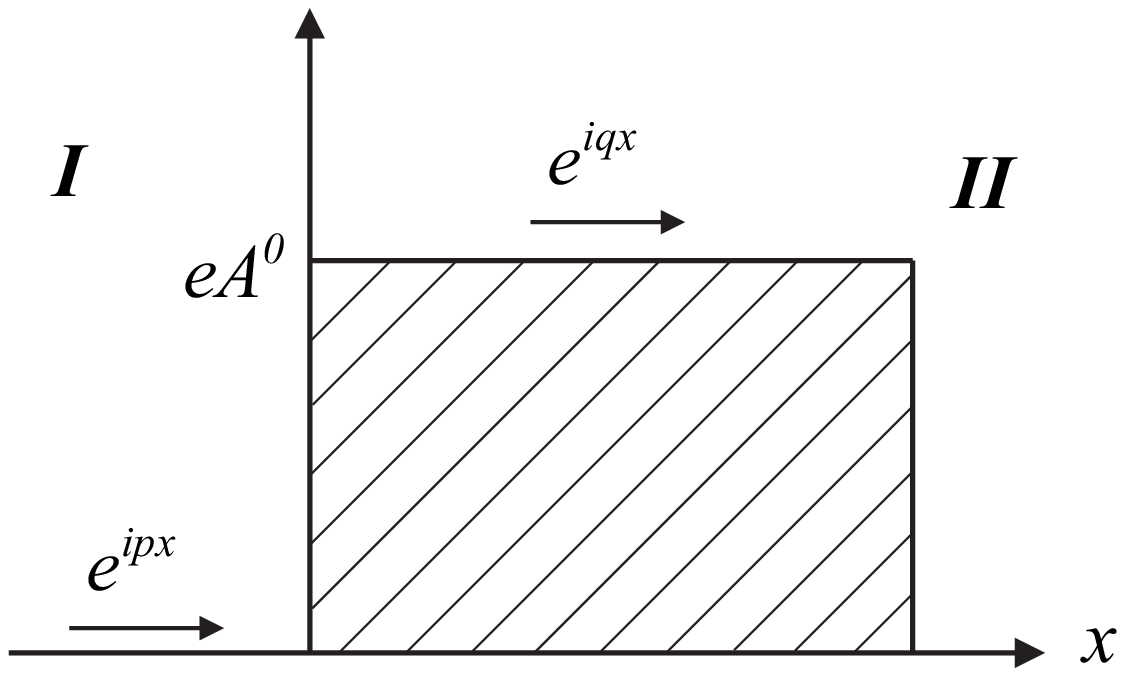}}
	%\caption{The physical vacuum of fermion equations with spinor functions.}
	\label{ris:Appendix}
\end{figure}

%%%%%%%%%%%%%%%%%%%%%%%%%%%%%%%%%%%%%%%%%%%%%%%%%%%%%%%%%%%%%%%%%%%%%%%%%

In domain I, $E=\beta \sqrt {m^{2}+p^{2}} $.

In domain II, $E-eA^{0}=\beta \sqrt {m^{2}+q^{2}} $.

For the FW-representation, there is no relationship between the solutions of 
the FW equation with different signs of $\beta $. In 
compliance with the Figure, in domain I, $\beta =1,E>0$. Hence, $\beta 
=1,E-eA^{0}\ge 0$ in domain II as well. For imaginary $q$, $\left| q 
\right|^{2}\le m^{2}$ should be fulfilled.

The Klein paradox is not available. The similar examination can also be 
performed for (QED) $_{KG}$.

%%%%%%%%%%%%%%%%%%%%%%%%%%%%%%%%%%%%%%%%%%%%%%%%%%%%%%%%%%%%%%%%%%%%%%%%%%%%

%\begin{thebibliography}{000} %for 3 digits

\end{document}